# Multiple optical traps from a single laser beam using a mechanical element

**J.A. Dharmadhikari, A.K. Dharmadhikari, and D. Mathur[*]**

Tata Institute of Fundamental Research, 1 Homi Bhabha Road, Mumbai 400 005, India

Abstract:

The use of a wire mesh facilitates creation of multiple optical traps for manipulation of small micron or sub-micron particles. Such an array of optical traps can be easily controlled. The trap that is formed in this manner is a continuous trap; it obviates the need to time share a laser beam among a set of positions, as is presently done in conventional multiple traps.

[*] e-mail- atmol1@tifr.res.in

Optical tweezers work on the principle of creating an intensity gradient by strongly focusing a laser beam. The resulting intensity gradient produces a field gradient that helps to trap a dielectric particle. Upon moving the physical focus point, the trapped particle also moves. The key feature of optical tweezers is that the steep intensity gradient induces a force whose magnitude is made stronger than the forces that are imparted by the scattering of the laser beam from the microscopic particles. The magnitude of such gradient forces can be readily and precisely controlled by varying the laser intensity. Particle sizes ranging from tens of nanometers to a few micrometers can be trapped and manipulated using single optical tweezers.[1,2] Although optical tweezers have found numerous and diverse applications, the fact that only a single particle can be trapped with one laser beam is a constraint that is being faced in many potential applications in the field of electronics, photonics, and biology where simultaneous spatial manipulation of a number of microscopic particles would be of considerable utility.[3]

Hitherto-existing multiple optical traps [3] rely on (i) the use of multiple laser beams, or (ii) on splitting a single laser beam in two parts using a polarizing beam splitter and then combining the two beams at the aperture of the objective, or (iii) time-sharing a single beam at different locations using an acousto-optic modulator, or (iv) making use of diffractive optical elements such as computer generated holograms which split the input beam into a pre-selected desired pattern.[4-7] Though the use of multiple laser beams appears to be the obvious choice to create multiple traps the cost factor and the associated difficulty of aligning more than a single beam through a 100X objective have proved to be serious experimental constraints.

The diffraction of monochromatic light from an array of apertures has been well known for several centuries. Such arrays, or wire meshes, constitute a simple diffractive element that localizes a single laser beam in specific locations. The possibility of regularizing spatial modulation of a femtosecond laser pulse has been recently demonstrated using a wire mesh[8,9] in the context of studies of the propagation properties of intense ultrashort laser light through condensed media. Here, we describe a very simple method based on the use of a metallic wire mesh to create multiple optical traps by diffracting an incident continuous wave laser beam. The wire mesh creates a diffraction pattern in a 2-dimensional plane perpendicular to the laser propagation direction, and the pattern is reproduced at specific distances that are obtained from Fresnel's diffraction law. Such a diffraction pattern, when focused using a microscope objective, produces multiple arrays, each of which constitutes an optical trap. The spacing between each trap can be altered by varying the wire mesh size. The multiple traps can be translated and rotated by fixing the wire mesh on a stage whose motion can be precisely controlled.

Figure 1 is a schematic representation of our experimental setup. The details of the trap are described elsewhere.[10] In brief a 1W, Nd-YVO$_4$ laser beam (1 mm diameter) is passed through beam expander so as to produce a 10 mm diameter parallel beam. This beam is then allowed to pass through two lenses (L1, L2) that form a telescopic arrangement, and finally through a 45° mirror on to a 100X microscope objective. A wire mesh of dimension 175 μm × 175 μm, with wire thickness of 95 μm and 50% transmission, is placed in the path of the beam near L1. The tightly focused beams that result from diffraction are used to trap and manipulate 1 μm diameter polystyrene beads. Altering the distance between the mesh and L1 changes the distance between adjacent traps. The strength of each trap is altered by changing the incident laser power. The reflected light from the sample is collected by a CCD camera and is interfaced to a computer through an image grabbing card to record real-time trapping events.

The diffraction pattern obtained from a mesh can be understood by considering a rectangular array of apertures with dimensions $b_x$ and $b_y$ that are separated by widths, $w_x$ and $w_y$, lying in the x-y plane. A plane monochromatic wave traveling along the z-direction incident normally on the above mentioned rectangular array will diffract the light. The intensity of light that is diffracted at angle $\alpha$ with respect to the z-axis in the x-y plane is given by the Fraunhofer relation[11]

$$I_x(\alpha) = \left( \frac{b_x \sin \beta_x}{\beta_x} \frac{\sin(N\delta_x)}{\sin \delta_x} \right)^2, \qquad (1)$$

where $\beta_x = \left( b_x \pi / \lambda \right) \sin \alpha$ and $\delta_x = \left( p_x \pi / \lambda \right) \sin \alpha$, $N$ is the number of diffracting elements contributing to the optical field, $\lambda$ is the wavelength of light used, and $p_x = b_x + w_x$ is the pitch of the array. The first factor in eqn.(1) is the Fraunhofer diffraction pattern due to a single slit of width $b_x$ and the second term is the intensity distribution resulting from interference between the diffraction patterns produced by the array of $N$ such slits whose pitch is $p_x$. In case of a square mesh (of the type we use) the intensity due to diffraction along the x and y directions is the same ($I_x = I_y$). The diffracted intensity in case of an array of apertures not only depends on the size of aperture $b$ but also on the pitch of the array $p$. We also carried out experiments with different sizes of the mesh placed before L1. The spacing between the two consecutive traps reduces when the mesh size is increased.

For a given size of mesh the spacing between two consecutive traps can be altered by placing the mesh in between the telescopic arrangement and by changing the distance between L1 and the mesh. It is evident from Fig. 2 that as the distance between the mesh and L1 increases, the distance between adjacent traps reduces. This is because increase in the distance between L1 and the mesh reduces the size of the laser beam that is incident on the mesh. The mesh is thus subjected to a convergent beam of light rather than a parallel beam. The second lens, L2, collects the divergent light from the mesh and converts it back into a parallel beam. With reference to Fig. 2a, when the mesh is very close to L1 (~ 1 cm), adjacent traps are separated by 8 μm. When the mesh is 5 cm away from the lens this distance reduces to 6 μm (Fig. 2b). For the separation of 11 cm the distance between the two traps is made as small as 3.4 μm (Fig. 2d). It should be noted that the central spot corresponds to the zeroth order diffraction; the immediately adjacent spots correspond to first order diffraction. If the incident laser power is further increased, second order diffraction spots also become visible.

We note that the diffraction efficiency remains constant with distance between the mesh and the objective: as long as all of the individual traps are from the same diffraction order, there is no variation in trap efficiency.

In order to demonstrate multiple trapping we carried out experiments using 1 μm diameter polystyrene spheres suspended in distilled water. The trapping events were captured in a real time movie (see supplementary material) and then converted to snapshots. Figures 3 a-d are typical snapshots showing beads that are trapped at different locations. By increasing the concentration of the beads we were able to demonstrate the trapping of a larger number of beads (Figs. 3 c-d). Now we see trapping at the location of first order diffraction and also at the second order diffraction. The trapped beads could be linearly translated using a precision translation stage, as is seen in the movie clip (see supplementary material).

All the available techniques of creating two or more optical traps have some advantages and disadvantages. In case of the acousto-optic deflector (AOD) even though one can control two traps independently, the traps are not continuous, their intermittency depending on the frequency of the AOD. Traps using diffractive optical elements have a fixed configuration for a static hologram. In order to change the trap configuration one has to change the diffractive optical element. Only in the case of a dynamic hologram can the configuration be easily altered; such traps are fully automated [12,13] but the associated degree of complexity (and cost) is much higher than with the trap described in this work which turns out to be a remarkably simple and inexpensive way to manipulate micron size particles.

In summary, we have demonstrated a simple technique to generate multiple traps from a single laser beam by using a wire mesh. Multiple trapping of a number of polystyrene beads is shown by way of illustration. There are several attractive features of our proposal: (i) the separation between adjacent trapping centers is easily controlled by changing the position of the mesh; (ii) the diffraction efficiency, hence the trapping efficiency, can be readily enhanced by using a mesh made of thinner wire (thereby increasing the mesh's transmission efficiency); (iii) very high incident laser power can be readily used; and (iv) variability of inter-spot spacing and spot size is readily accessible.

JAD thanks the Homi Bhabha Fellowship Council for financial support.

Figure Captions

1. Experimental setup to create multiple traps with a wire mesh (175 μm × 175 μm, wire thickness 95 μm)

2. Photographs showing the variation of the distance between adjacent traps accomplished by varying the distance between the wire mesh and the lens L1: a) 1 cm, b) 5 cm, c) 7 cm and d) 11 cm. The corresponding distances between adjacent traps were a) 8 μm, b) 6 μm, c) 5 μm, and d) 3.4 μm.

3. Snapshots showing trapping of 1 μm size polystyrene beads in multiple traps. The distance between the lens and the wire mesh is kept constant at 5 cm. Beads are trapped at different locations: trapping of 3 beads (a) and 4 beads (b) at zeroth and first order diffraction locations, trapping of 5 beads (c) and 6 beads (d) at zeroth, first, and second order diffraction locations.

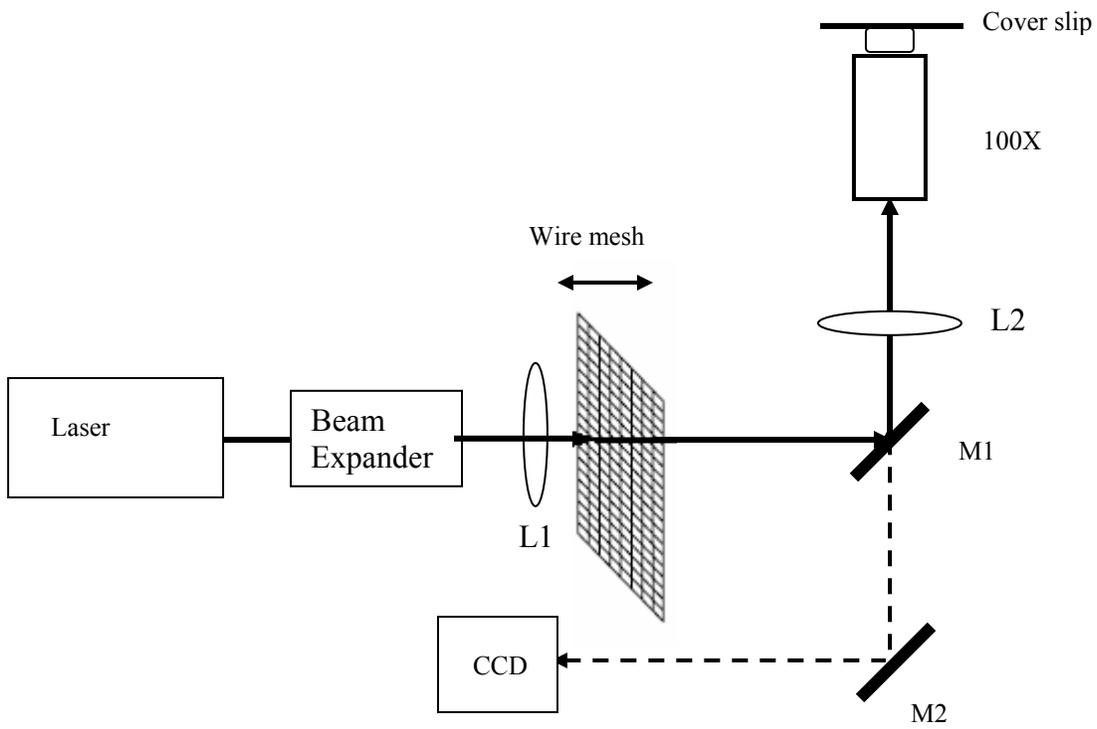

Fig. 1

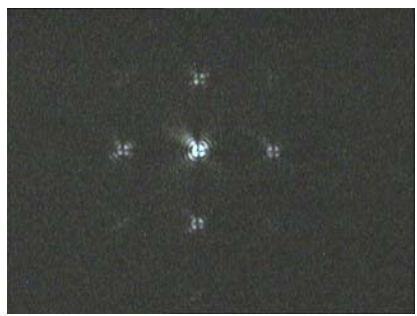 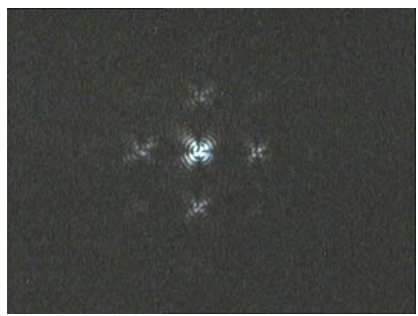

a) 1 cm　　　　　　　　　　　　　b) 5 cm

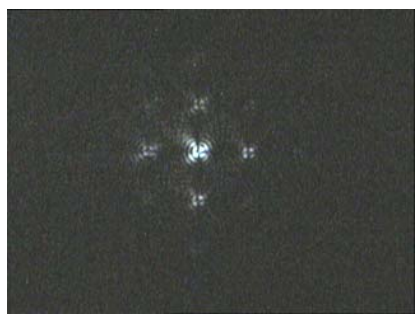 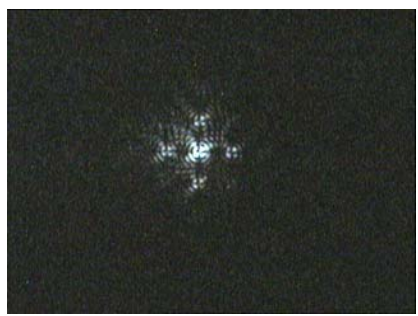

c) 7 cm　　　　　　　　　　　　　d) 11 cm

Fig. 2

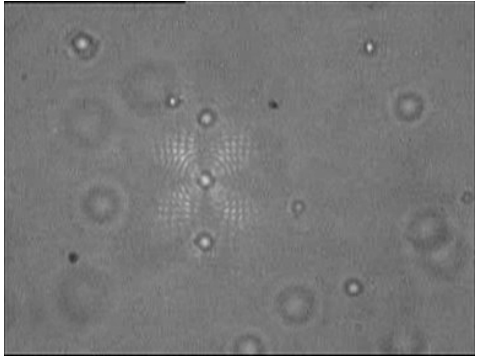 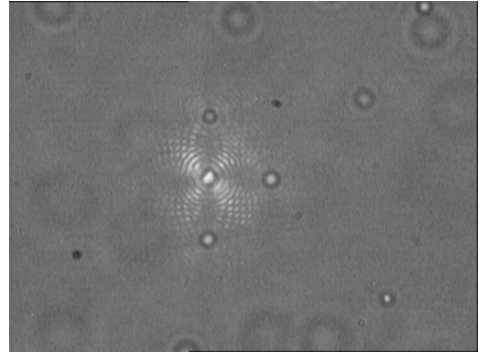

(a) (b)

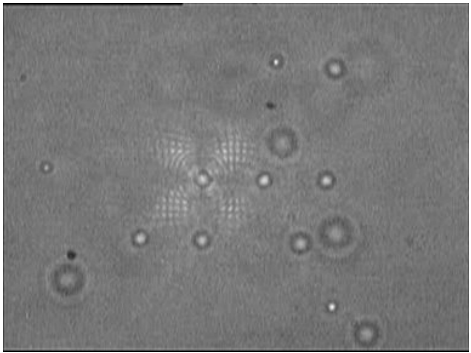 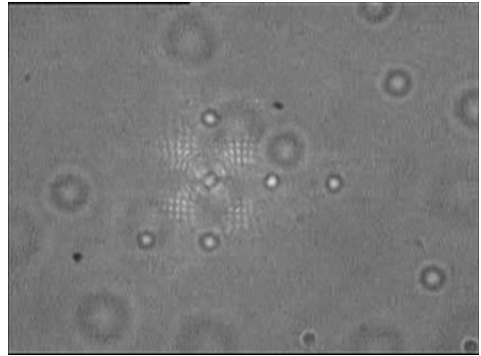

(c) (d)

Fig. 3